\documentclass[letterpaper, 10 pt, conference]{ieeeconf}  

\IEEEoverridecommandlockouts                              

\usepackage{cite}
\usepackage{amsmath,amssymb,amsfonts}
\usepackage{algorithm}
\usepackage{graphicx}
\usepackage{textcomp}
\usepackage{xcolor}
\usepackage{algpseudocode}
\newtheorem{theorem}{Theorem}
\newtheorem{lemma}[theorem]{Lemma}
\newtheorem{definition}{Definition}
\def\BibTeX{{\rm B\kern-.05em{\sc i\kern-.025em b}\kern-.08em
    T\kern-.1667em\lower.7ex\hbox{E}\kern-.125emX}}

\makeatletter
\newcommand\fs@betterruled{%
  \def\@fs@cfont{\bfseries}\let\@fs@capt\floatc@ruled
  \def\@fs@pre{\vspace*{5pt}\hrule height.8pt depth0pt \kern2pt}%
  \def\@fs@post{\kern2pt\hrule\relax}%
  \def\@fs@mid{\kern2pt\hrule\kern2pt}%
  \let\@fs@iftopcapt\iftrue}
\floatstyle{betterruled}
\restylefloat{algorithm}
\makeatother

\title{\LARGE \bf
Algorithmic (Semi-)Conjugacy via Koopman Operator Theory
}

\author{William T. Redman$^{1, 2}$, Maria Fonoberova$^{2}$, Ryan Mohr$^{2}$, Ioannis G. Kevrekidis$^{3}$, and Igor Mezić$^{2, 4}$
\thanks{$^{1}$W.T.R. is with the Interdepartmental Graduate Program in Dynamical Neuroscience, University of California, Santa Barbara.}%
\thanks{$^{2}$W.T.R., M.F., R.M., and I.M. are with AIMdyn Inc., Santa Barbara.}
\thanks{$^{3}$ I.G.K. is with the Departments of Chemical and Biomolecular Engineering, and of Mathematics and Statistics, The Johns Hopkins University.}
\thanks{$^{4}$ I.M. is with the Departments of Mechanical Engineering and of Mathematics, University of California, Santa Barbara.}
}

\begin{document}

\maketitle
\thispagestyle{empty}
\pagestyle{empty}

\begin{abstract}

Iterative algorithms are of utmost importance in decision and control. With an ever growing number of algorithms being developed, distributed, and proprietarized, there is a similarly growing need for methods that can provide classification and comparison. By viewing iterative algorithms as discrete-time dynamical systems, we leverage Koopman operator theory to identify (semi-)conjugacies between algorithms using their spectral properties. This provides a general framework with which to classify and compare algorithms. 

\end{abstract}

\section{Introduction}

For many problems that are commonly faced in control and decision settings, a variety of numerical algorithms exist to find approximate solutions. For instance, ordinary differential equations can be solved with either the forward or the backward Euler method. Roots to polynomial functions can be found with either the Newton-Raphson or Secant method. Deep neural networks can be optimized with stochastic gradient descent using either a fixed or an adaptive learning rate. In each case, differences in numerical stability, usage of computational resources, and speed, among other factors, must be taken into consideration, in order to decide which method to employ.

A well known result within the field of numerical analysis is that many of these seemingly different algorithms are, in fact, equivalent. That is, the output of one algorithm can be exactly mapped to the output of another. In its simplest form, such equivalency between algorithms involves the same number of variables and operations, but different values of free parameters, making an appropriate choice of initial conditions lead to the same output. A sufficiently straightforward example of such an equivalence could be identified by looking at the underlying equations. However, in more subtle scenarios, two algorithms can be equivalent, but take on different looking forms, making an identification challenging. The ability to recognize equivalency is important in assessing the novelty of  proposed algorithms \cite{zhao2021automatic}. Additionally, classifying algorithms by their equivalent counterparts (i.e. defining equivalence classes) provides a way in which to better analyze them.

Recent work to automatically detect equivalencies has taken a control theoretic perspective \cite{zhao2021automatic}. By viewing iterative algorithms as linear control systems \cite{lessard2016analysis}, and evaluating their corresponding transfer functions, it was possible to develop an analytic method that could directly identify equivalent algorithms. While broadly successful, such an approach has several limitations. First, it was only able to identify linear mappings between algorithms, restricting the class of equivalencies that could be identified. Second, several different definitions of equivalence had to be introduced in order to capture the equivalencies of the different algorithms that were studied. Third, it was necessary to have access to the equations underlying the iterative algorithms. In cases where the algorithms are proprietary, this may not be available. Lastly, the control theoretic framework adds an additional layer of complexity that is not strictly necessary.

We study the equivalence of algorithms from a related, dynamical systems theoretic perspective. Dynamical systems have a long history of inter-relatedness with algorithms \cite{stuart1996dynamical, chu2008linear, sahai2020dynamical}, and recent developments in Koopman operator theory \cite{Koopman1931, Koopman1932, mezic2005spectral, budisic2012applied} have been used to optimize and analyze algorithms \cite{dietrich2020algorithms}, especially within the field of machine learning  \cite{dogra2020optimizing, tano2020accelerating, manojlovic2020applications, naiman2021sequence, mohr2021predicting, redman2021operator}. Here, we show that the different types of equivalence established by Zhao et al. (2021) \cite{zhao2021automatic} naturally fall under the idea of (semi-)conjugacy, which can be identified from spectral properties of the Koopman operator. Using several of the illustrative examples from Zhao et al. (2021), we find that different algorithms that are equivalent by: 1) linear, invertible mappings; 2) linear, embedded mappings; 3) nonlinear, invertible mappings; and 4) shifts of output values, can all be detected by signatures in their Koopman spectra. Taken together, our work provides evidence that Koopman operator theory is a general approach for studying algorithmic equivalence. 

The remainder of the paper is organized as follows. We begin by discussing how iterative algorithms can be viewed as discrete-time dynamical systems (Sec. \ref{subsec:iterative algos as dynamical systems}) and review basics of Koopman operator theory (Sec. \ref{subsec:Koopman mode decomposition}). In particular, we highlight the importance of the Koopman mode decomposition. We proceed to define the notions of (semi-)conjugacy, and discuss results that allow the Koopman spectrum to be used in identifying (semi-)conjugate systems (Sec. \ref{subsec:Conjugacy}). In Sec. \ref{sec:results}, we show that the theory outlined in Sec. \ref{sec:KOT} can be used to identify equivalences in various example algorithms. We conclude in Sec. \ref{sec:discussion}. 

\section{Koopman Operator Theory}
\label{sec:KOT}

\subsection{Iterative algorithms as dynamical systems}
\label{subsec:iterative algos as dynamical systems}
The idea central to our framework is that iterative algorithms can be viewed as discrete-time dynamical systems. To see this, consider an algorithm, $A:X \rightarrow X$, whose state space is $X \subseteq \mathbb{R}^d$, for some $d \in \mathbb{N}$. Starting from an initial state $x_0 \in X$, each iterative application of $A$ evolves the state-space input by
\begin{equation}
\label{eq:iterative algo finite time DS}
    x_{k+1} = Ax_k = A^{k + 1}x_0.
\end{equation}
Comparing Eq. \ref{eq:iterative algo finite time DS} to the classical perspective of dynamical systems, we see that $A$ acts as a dynamical map, with the number of iterations, $k \in \mathbb{N}$, acting as ``time''. The resulting sequence, $\{x_k\}_{k=1}^N$, is a trajectory through state-space. The existence of geometric objects that are studied in dynamical systems theory, such fixed points, limit cycles, quasi-periodic attractors, etc. depends on the precise nature of $A$ and $X$. Indeed, the same $A$, applied to different domains of $X$, can have different properties. However, in order for an algorithm to be of general practical utility, it is necessary for it to converge, and converge in a finite number of iterations. Therefore, for the algorithms we study in this paper, we will assume that there is a large region of state space, $M \subseteq X$, where $A$ evolves any initial condition $x_0 \in M$ to within $\varepsilon$ of the fixed point $x^*$, in at most $N_{\text{max}} \in \mathbb{N}$ iterations. 

However, the key idea in this paper, the comparison of dynamical systems based on the spectrum of the associated Koopman operator, is applicable to a much broader set of systems with point, limit cycle, toroidal, and even chaotic attractors \cite{mezic2016comparison,mezic2020spectrum}.

\subsection{Koopman mode decomposition}
\label{subsec:Koopman mode decomposition}
While the discrete-time dynamical systems view of algorithms outlined in Sec. \ref{subsec:iterative algos as dynamical systems} motivates the use of dynamical systems theory, the standard tools are difficult to leverage when the algorithms are nonlinear or when the equivalence between algorithms is nonlinear. Additionally, they are impossible to use when the equations underlying the algorithms are not known. Therefore, the course of action is not immediately apparent. Koopman operator theory, a data-driven dynamical systems theory \cite{Koopman1931, Koopman1932, mezic2005spectral, budisic2012applied} that is intimately related to the geometrical objects present in the classical theory \cite{mezic2005spectral,  mauroy2013iso, lan2013linearization, arbabi2017ergodic, mezic2020spectrum}, offers a way forward. 

The central object of interest in Koopman operator theory is the Koopman operator, $U$, an infinite dimensional linear operator that describes the time evolution of observables (i.e. functions of the underlying state space variables) that live in the functional space, $\mathcal{F}$. That is, after $t > 0$ amount of time, which can be continuous or discrete, the value of the observable $f \in \mathcal{F}$, which can be a scalar or a vector valued function, is given by
\begin{equation}
    \label{KO}
    U^t f(x_0) = f \left[ T^t(x_0) \right],
\end{equation}
where $T$ is the dynamical map evolving the system and $x_0 \in M$ is the initial condition in state space $M \subseteq X \in \mathbb{R}^d$. For the remainder of the paper, it will be assumed that $M$ is of finite dimension and that $\mathcal{F}$ is the suitably chosen space of  functions in which the spectral expansion exists \cite{mezic2020spectrum}.

The action of the Koopman operator on the observable $f$ can be decomposed as 
\begin{equation}
\label{Koopman mode decomposition}
     U f(x_0) = \sum_{r = 0}^\infty \lambda_r \phi_r(x_0) v_r,
\end{equation}
where the $\phi_r$ are eigenfunctions of $U$, with $\lambda_r \in \mathbb{C}$ as their eigenvalues and $v_r$ as their Koopman modes \cite{mezic2005spectral}. For systems with chaotic or shear dynamics, there is an additional term in Eq. \ref{Koopman mode decomposition} arising from the continuous part of the spectrum \cite{mezic2005spectral}. As the algorithms we study are not expected to have such dynamics, for the remainder of this paper it will be assumed that the dynamical systems we are considering are such that the Koopman operator only has a point spectrum. 

Spectrally decomposing the action of the Koopman operator is powerful because, for a discrete-time dynamical system, the value of $f$ at time step, $k \in \mathbb{N}$, is given simply by
\begin{equation}
    \label{Koopman mode decomposition time}
     f\left[ T^k(x_0) \right]  = U^k f(x_0) = \sum_{r = 0}^\infty \lambda_r^k \phi_r(x_0) v_r.
\end{equation}

From Eq. \ref{Koopman mode decomposition time}, we see that the dynamics of the system in the directions $v_r$, scaled by $\phi_r(x_0)$, are given by the magnitude of the corresponding $\lambda_r$. Assuming that $|\lambda_r| \leq 1$ for all $r$, finding the long time behavior of $f$ amounts to considering only the $\phi_r(x_0)v_r$ whose $|\lambda_r| \approx 1$. 
 
While the number of triplets $(v_r, \lambda_r, \phi_r)$ needed to fully capture the action of $U$ is, in principle, infinite, in many applied settings it has been found that a finite number, $R \in \mathbb{N}$, of them allows for a good approximation \cite{budisic2012applied}. Namely, for a generic $n$-dimensional system that is stable in the basin of attraction to a fixed point $x^*$, there is the set of principal eigenvalues $\boldsymbol \lambda =(\lambda_1,  \lambda_2,..., \lambda_R)$, with real part less than or equal to $1$ and such that $|\lambda_1|<|\lambda_2|<...<|\lambda_R|$. All the other eigenvalues are obtained by
\begin{equation}
    \lambda^\mathbf{k}=\lambda_1^{k_1}\cdot \lambda_2^{k_2}...\cdot\lambda_n^{k_R}, 
\end{equation}
where $\mathbf{k}=(k_1,...,k_R)\in \mathbb{N}^R$. Thus, we can often select a finite number, $R$, of eigenvalues that dominate the spectral expansion after a sufficient number of iterates, as their magnitudes are closer to $1$. That is, 
\begin{equation}
    \label{Koopman mode decomposition finite}
    U^k f(x_0) \approx \sum_{r = 0}^{R - 1} \lambda_r^k \phi_r(x_0) v_r.
\end{equation}

Many numerical methods exist to compute the Koopman mode decomposition, which have allowed it to be applied to complex problems across a wide range of scientific domains. The most popular method is dynamic mode decomposition (DMD) \cite{schmid2010dmd, rowley2009dmd, tu2014dmd}, and its variants \cite{chen2012variants, williams2015edmd, arbabi2017ergodic}. Here we make use of DMD and Extended-DMD \cite{williams2015edmd} because of their ubiquity.

\subsection{(Semi-)Conjugacy}
\label{subsec:Conjugacy}
As discussed in the previous section, in a suitably chosen functional space, all eigenvalues of the Koopman operator are obtained from the $R$ principal ones. The key idea in comparing algorithms then becomes one of comparing principal eigenvalues. This is justified by utilizing the notion of \newline (semi-)conjugacy. 
\begin{definition} Let $T:M\rightarrow M$ and $S:N\rightarrow N$ be two discrete-time dynamical systems on sets $M\subset \mathbb{R}^m$ and $N\subset \mathbb{R}^n$, with the associated Koopman operators $U^T$ and $U^S$. They are conjugate provided $m=n$ and there exists a smooth invertible mapping $h:M\rightarrow K$, such that $h\circ T=S\circ h$. In other words, $U^Th=S\circ h$. If $n<m$ and $h$ is smooth but not invertible, then the mapping is called semi-conjugate.
\end{definition}

Clearly, if $T$ and $S$ are semi-conjugate, then $h(x^*)$ is a fixed point of $S$ if and only if $x^*$ is a fixed point of $T$. We have the following Lemma.
\begin{lemma} If $T$ and $S$ are conjugate and $T$ is stable in the basin of attraction to $x^*$, then they have the same principal eigenvalues. If $T$ and $S$ are semi-conjugate, then the set of principal eigenvalues of $U^S$ is a subset of the set of principal eigenvalues of $U^T$. 
\end{lemma}

{\it Proof:} Let $\lambda,\phi$ be an eigenvalue, eigenfunction pair of $U^S$. Then
\begin{equation}
\begin{split}
   \lambda \phi[h(x)] = U^S \phi[h(x)] &=  \phi\left(S[h(x)]\right)=...\\ ...=\phi\left(h[T(x)]\right)&=U^T\phi(h). 
\end{split}
\end{equation}
Thus, $\lambda$ is an eigenvalue of $U^T$, associated with the eigenfunction $\phi\circ h$. The converse to Lemma 1, that is, if $U^T$ and $U^S$ have discrete spectra that are equivalent, then T and S are conjugate, can also be shown to be true \cite{mezic2016comparison}.

\section{Results}
\label{sec:results}
Having developed a dynamical systems framework for studying algorithmic equivalence, via comparing the spectra of the Koopman operators associated with the algorithms being applied to a given problem (Sec. \ref{sec:KOT}), we tested whether standard numerical methods could properly identify instances of (semi-)conjugacy. To do this, we made use of several of the illustrative examples that were presented by Zhao et al. (2021), examining equivalencies defined by various types of mappings (i.e. linear/nonlinear, invertible/embedded) \cite{zhao2021automatic}. In each case, we find that implementations of Koopman mode decomposition can identify the underlying conjugacy, and thus, provide a general framework for probing algorithmic equivalency.

\subsection{Equivalence by linear, invertible mappings}
We begin by examining the toy Algorithms \ref{algo: 1} and \ref{algo: 2} \cite{zhao2021automatic}. Each algorithm's behavior is determined by the choice of the function $f$, as $\nabla f$ is an operation that occurs in both algorithms. For any $f$, there exists an invertible, linear mapping between the two algorithms' outputs, such that the two sequences are equivalent, assuming the initial conditions have been properly set. Namely,
\begin{equation}
\label{eq:algoriths 1 and 2 equivalence}
    \begin{split}
        \xi_1^k = 2x_1^k - x_2^k \\
        \xi_2^k = -x_1^k + x_2^k
    \end{split}
\end{equation}
defines a mapping between the output of Algorithm \ref{algo: 1} ($x_i^k$) and the output of Algorithm \ref{algo: 2} ($\xi_i^k$). In this sense, the two algorithms are ``oracle equivalent'' \cite{zhao2021automatic}, where $\nabla f$ is considered an oracle. 

\begin{algorithm}
\caption{}
\label{algo: 1}
\begin{algorithmic}
\For{$k = 0, 1, 2, ..., K$} 
\State $x_1^{k + 1} = 2x_1^k - x_2^k - \frac{1}{10}\nabla f(2x_1^k - x_2^k)$ 
\State $x_2^{k + 1} = x_1^k$
\EndFor
\end{algorithmic}
\end{algorithm}

\begin{algorithm}
\caption{}
\label{algo: 2}
\begin{algorithmic}
\For{$k = 0, 1, 2, ..., K$} 
\State $\xi_1^{k + 1} = \xi_1^k - \xi_2^k - \frac{1}{5}\nabla f(\xi_1^k)$ 
\State $\xi_2^{k + 1} = \xi_2^k + \frac{1}{10}\nabla f(\xi_1^k)$
\EndFor
\end{algorithmic}
\end{algorithm}

\begin{figure}
    \centering
    \vspace{2mm}
    \includegraphics[width = 0.47\textwidth]{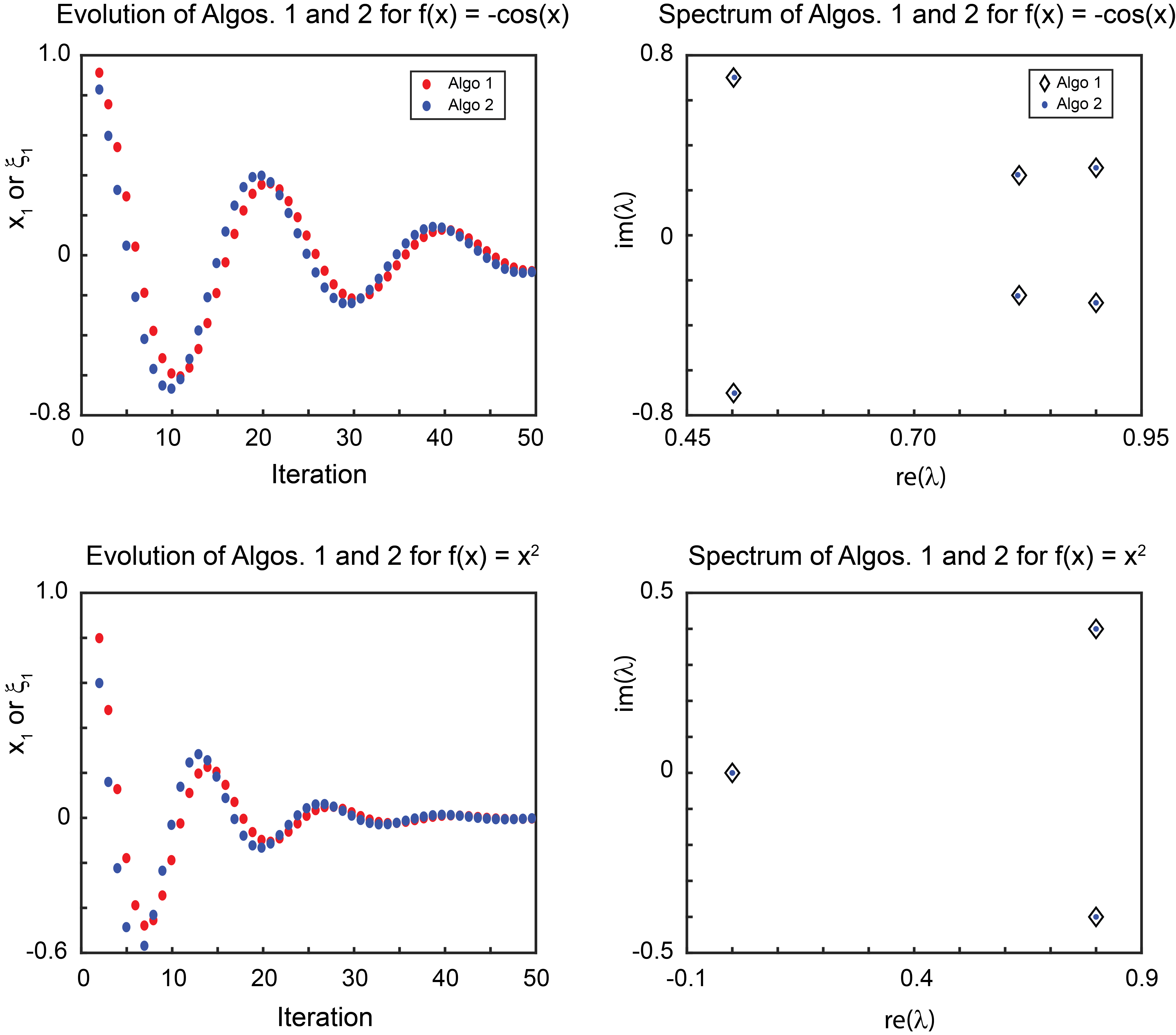}
    \caption{\textbf{Conjugacy of Algorithms \ref{algo: 1} and \ref{algo: 2}.} Left column, trajectory of state variables $x_1$ and $\xi_1$. Right column, corresponding Koopman spectra. Top row, $f = -\cos(x)$. Bottom row, $f = x^2$. The nearly identical spectra correctly illustrates that the two algorithms are conjugate.}
    \label{fig:Algo Linear Invertible Map}
\end{figure}

We examine the Koopman spectra of the two algorithms applied to pairs of initial conditions, related via Eq. \ref{eq:algoriths 1 and 2 equivalence}, for different choices of the function $f$. Examples of $f(x) = x^2$ and $f(x) = -\cos(x)$ are shown in Fig. \ref{fig:Algo Linear Invertible Map}.

In both cases, we find that the Koopman spectra are nearly exactly overlapping. Thus, as predicted from the theory discussed in Sec. \ref{sec:KOT}, conjugacy via linear, invertible mapping can be identified by the two algorithms having the same Koopman eigenvalues.

While Algorithms \ref{algo: 1} and \ref{algo: 2} are conjugate by Eq. \ref{eq:algoriths 1 and 2 equivalence} for both choices of $f(x)$ we considered, the nature of this conjugacy differs. When $f(x) = x^2$, the two algorithms are linear, as $\nabla f = 2x$, and the matrices governing their dynamics have the same eigenvalues, corresponding to the same single attractor. Therefore, Algorithms \ref{algo: 1} and \ref{algo: 2} are \textit{globally conjugate}, as any two initial conditions will evolve, with the same dynamics, to $(0, 0)$. 

In contrast, when $f(x) = -\cos(x)$, the algorithms are no longer linear, as $\nabla f = \sin(x)$, and there exist multiple attractors. For instance, when $2x_1^0 - x_2^0$ is small, Algorithm \ref{algo: 1} converges to $(0, 0)$, but when $2x_1^0 - x_2^0$ is large, $x_1^k$ and $x_2^k$ grow approximately linearly. A similar transition occurs for Algorithm \ref{algo: 2}. Therefore, Algorithms \ref{algo: 1} and \ref{algo: 2} are \textit{locally conjugate}, as initializing them differently can lead to trajectories in different dynamical regimes. 

\begin{figure}
    \centering
    \includegraphics[width = 0.49\textwidth]{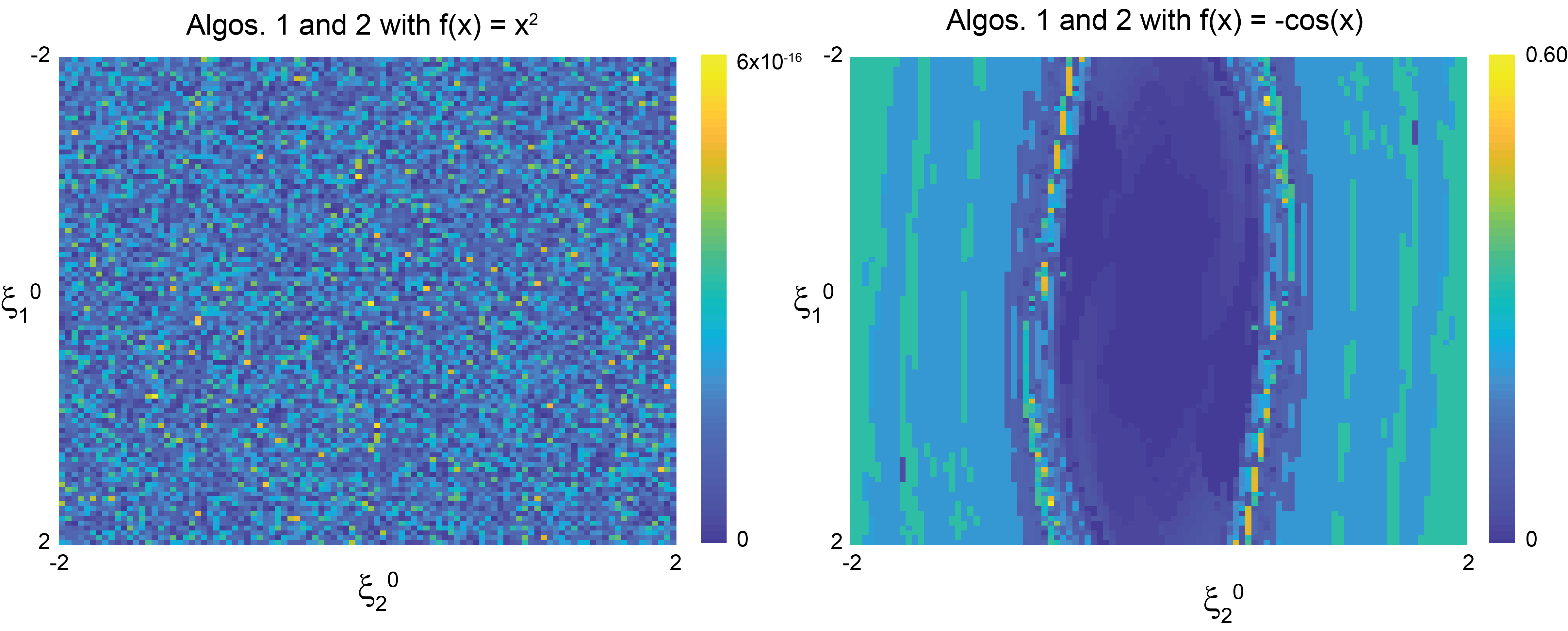}
    \caption{\textbf{Global and local conjugacy.} Wasserstein distance of the Koopman spectra when the initial condition for Algorithm \ref{algo: 1} is fixed at $x_1^0 = x_2^0 = 0.1$ and the initial condition of Algorithm \ref{algo: 2}, $\xi_1^0$ and $\xi_2^0$, is sampled over the range $[-2, 2]\times[-2, 2]$. Left, the Koopman spectra are nearly identical for every pair of initial conditions when $f(x) = x^2$. Right, the Koopman spectra can be similar or different when $f(x) = -\cos(x)$, depending on the initial conditions. }
    \label{fig:Wasserstein distance}
\end{figure}

To illustrate this, and demonstrate that Koopman operator theory can identify such differences in the behavior of the conjugacies, we fix $x_1^0 = x_2^0 = 0.1$ (making $2x_1^0 - x_2^0 = 0$ small), and vary $(\xi^0_1, \xi^0_2)$ over the interval $[-2, 2]\times[-2, 2]$. For each pair of $(\xi^0_1, \xi^0_2)$, we evolve Algorithm \ref{algo: 2}, approximate the Koopman spectrum, and compute the Wasserstein distance between it and the spectrum found when applying Algorithm \ref{algo: 1} to $(x_1^0, x_2^0) = (0.1, 0.1)$. The Wasserstein distance is a metric originally developed in the context of optimal transport, and provides a sense of how far apart two distributions are. As expected, when $f(x) = x^2$, all $(\xi^0_1, \xi^0_2)$ we evaluated led to the same dynamics of Algorithm \ref{algo: 2} as compared to Algorithm \ref{algo: 1}, with Koopman eigenvalues that were separated by less than $10^{-15}$ (Fig. \ref{fig:Wasserstein distance}, left). However, when $f(x) = -\cos(x)$, there is a region where the Wasserstein distance increases, as the dynamics of Algorithm \ref{algo: 2} transition to a different attractor (Fig. \ref{fig:Wasserstein distance}, right). 

These results highlight the fact that our Koopman framework can provide a view on the relationship between two algorithms that is broader than the binary classification of whether they are conjugate or not. Given that the nature of equivalence is an important factor to take into account when classifying and comparing algorithms, this is a useful feature. 

\subsection{Equivalence by linear, embedded mappings}

We next consider Algorithms \ref{algo: 3} and \ref{algo: 4} \cite{zhao2021automatic}. There exists a linear, embedded mapping between their outputs. Namely,

\begin{equation}
\label{eq:algoriths 3 and 4 equivalence}
    \begin{split}
        \xi^k = -x_1^k + 2x_2^k \\
    \end{split}
\end{equation}
defines a mapping of the two-dimensional output of Algorithm \ref{algo: 3} ($x_i^k$) to the one-dimensional output of Algorithm \ref{algo: 4} ($\xi^k$).

\begin{algorithm}
\caption{}
\label{algo: 3}
\begin{algorithmic}
\For{$k = 0, 1, 2, ..., K$} 
\State $x_1^{k + 1} = 3 x_1^k - 2 x_2^k + \frac{1}{5}\nabla f(-x_1^k + 2 x_2^k)$ 
\State $x_2^{k + 1} = x_1^k$
\EndFor
\end{algorithmic}
\end{algorithm}

\begin{algorithm}
\caption{}
\label{algo: 4}
\begin{algorithmic}
\For{$k = 0, 1, 2, ..., K$} 
\State $\xi^{k + 1} = \xi^k - \frac{1}{5}\nabla f(\xi^k)$ 
\EndFor
\end{algorithmic}
\end{algorithm}

\begin{figure}
    \centering
    \includegraphics[width = 0.47\textwidth]{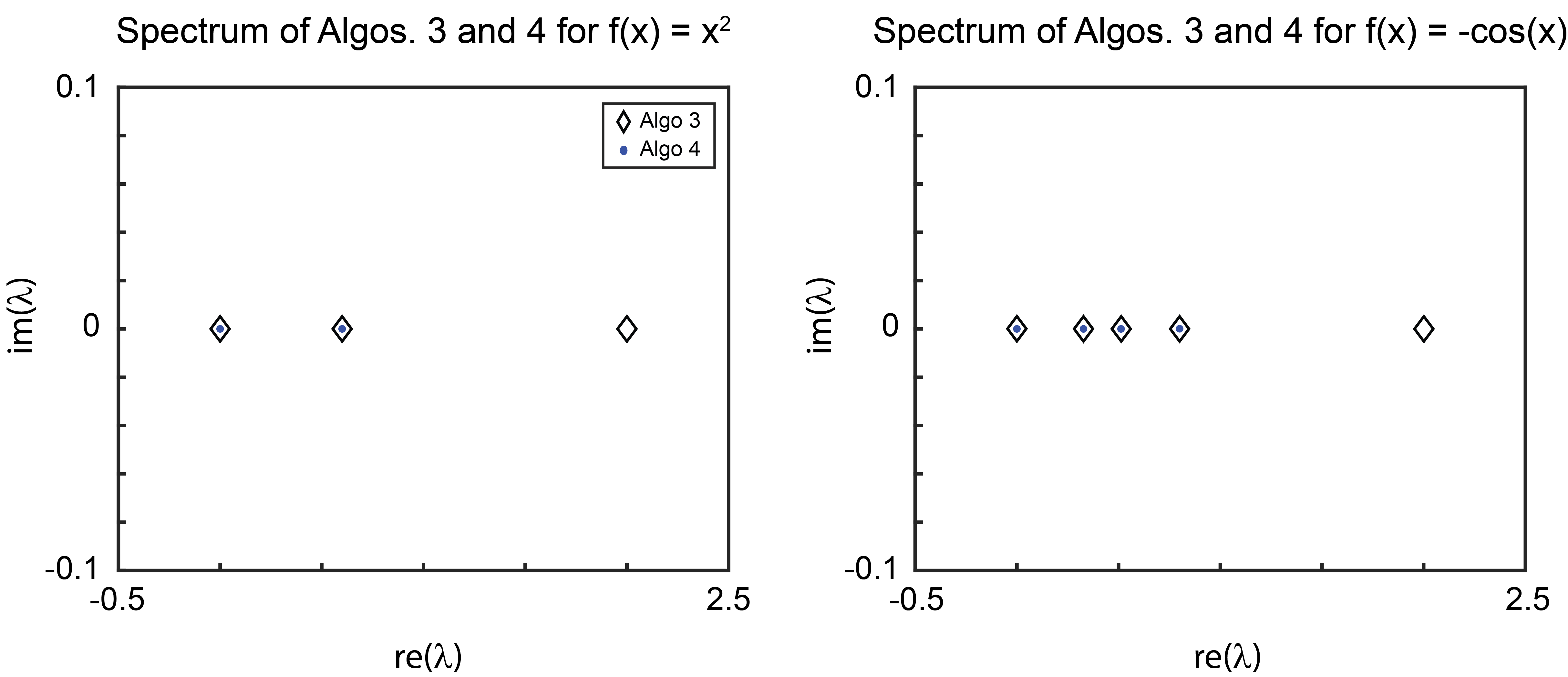}
    \caption{\textbf{Semi-conjugacy of Algorithms \ref{algo: 3} and \ref{algo: 4}.} The spectrum of Algorithm \ref{algo: 4} is a subset of the spectrum of Algorithm \ref{algo: 3}, correctly implying that the two systems are semi-conjugate, for both $f(x) = x^2$ (left) and $f(x) = -\cos(x)$.}
    \label{fig:Algo Linear Noninvertible}
\end{figure}

From the theory developed in Sec. \ref{sec:KOT}, we can recognize Eq. \ref{eq:algoriths 3 and 4 equivalence} as describing a semi-conjugacy. We therefore expect that the Koopman spectrum associated with the smaller system (Algorithm \ref{algo: 4}) should be a subset of the spectrum associated with the larger system (Algorithm \ref{algo: 3}). Indeed, we find this to be the case for $f(x) = x^2$ and $f(x) = -\cos(x)$ (Fig. \ref{fig:Algo Linear Noninvertible}). In both cases, the spectra have the same eigenvalues corresponding to decaying modes (i.e. $|\lambda| < 1$). However, the eigenvalue corresponding to the exponentially growing mode (i.e. $|\lambda| > 1$) is only present in the spectrum of Algorithm \ref{algo: 3}, which indeed matches the unbounded growth of $x_1^k$ and $x_2^k$. 

\subsection{Equivalence by nonlinear, invertible mappings}

We next tackle an equivalence noted by Zhao et al. (2021) that is given by a nonlinear, invertible mapping. This is a non-trivial problem to identify, and cannot be done within a linear control framework \cite{lessard2016analysis, zhao2021automatic}. 

\begin{figure}
    \centering
    \vspace{1.5mm}
    \includegraphics[width = 0.49\textwidth]{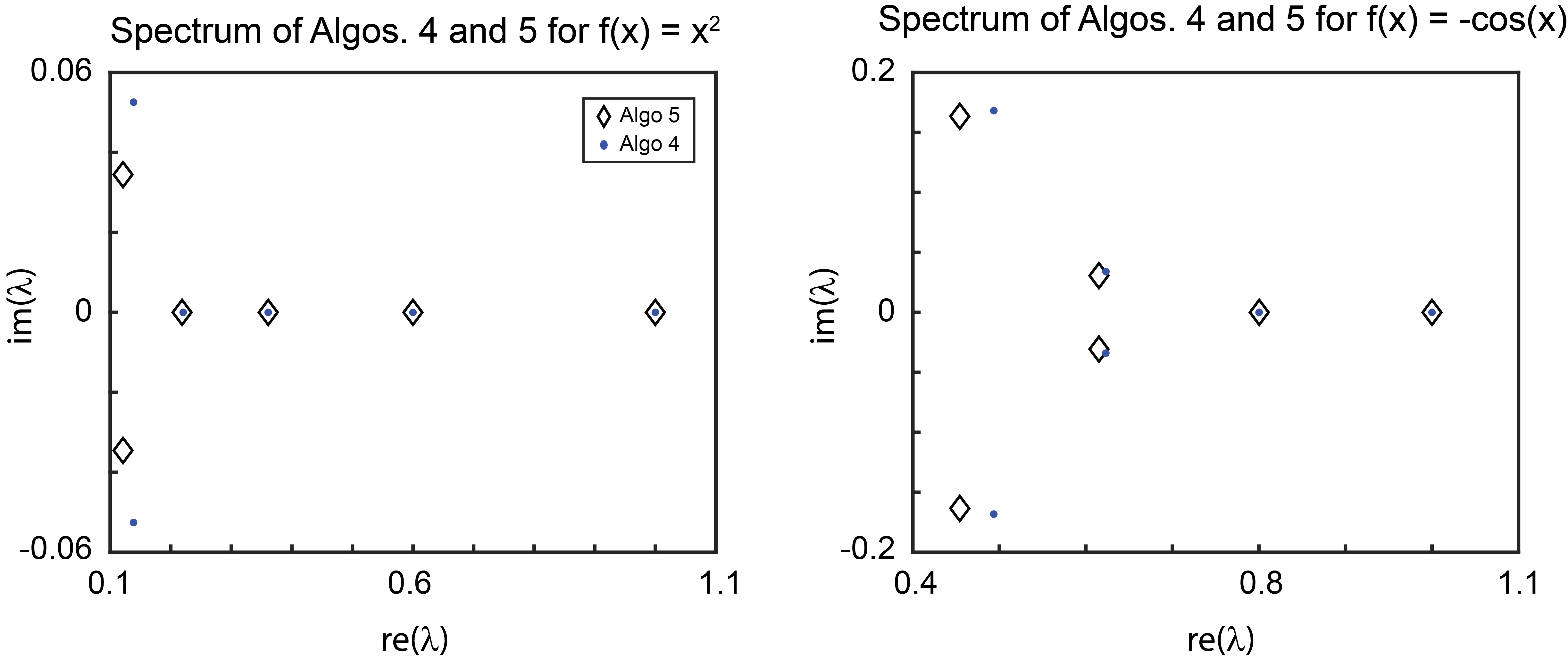}
    \caption{\textbf{Conjugacy of Algorithms \ref{algo: 4} and \ref{algo: 5}.} The nearly overlapping spectra, up to some differences in the smallest eigenvalues, correctly implies that Algorithms \ref{algo: 4} and \ref{algo: 5} are conjugate.}
    \label{fig:Algo Nonlinear Invertible Map}
\end{figure}

Algorithm \ref{algo: 5} is equivalent to Algorithm \ref{algo: 4} via,
\begin{equation}
    x^k = \exp(\xi^k).
\end{equation}
From the theory discussed in Sec. \ref{sec:KOT}, we expect that the Koopman spectra of such conjugate systems should be equivalent. We indeed find this to be true (Fig. \ref{fig:Algo Nonlinear Invertible Map}). 

\begin{algorithm}
\caption{}
\label{algo: 5}
\begin{algorithmic}
\For{$k = 0, 1, 2, ..., K$} 
\State $x^{k + 1} = x^k \exp(-\frac{1}{5}\nabla f(\log x^k))$ 
\EndFor
\end{algorithmic}
\end{algorithm}

We note that, since we used a simple numerical scheme to compute the Koopman mode decomposition on a small set of data points, we do not see a perfect overlap of the spectra. In particular, the smallest eigenvalues differ the most between the two. Given the plethora of robust, precise numerical methods, we imagine it would be possible to achieve better overlap. Regardless, the fact that the largest eigenvalues overlap suggests a clear connection between the two algorithms. Given that any numerical approach will find some non-zero difference, there will necessarily have to be some discretion practiced by the user. 

\subsection{Equivalence by shifting}

Because algorithms can perform the same sequence of operations, but do so in differing orders, Zhao et al. (2021) considered ``shift equivalence''. By permuting the transfer operators associated with each algorithm, it became possible to identify shift equivalent algorithms under the linear control framework \cite{zhao2021automatic}. However, this required some additional theory. We reasoned that our Koopman operator theoretic approach may be able to identify shift equivalent algorithms, without any new machinery. 

To explore this, we considered Algorithms \ref{algo : 6} and \ref{algo : 7} \cite{zhao2021automatic}. The two make use of the proximal operator, $\text{prox}_f$. To evaluate this operation numerically, we make use of the UNLocBoX package \cite{perraudin2014unlocbox}. The two algorithms are equivalent via 
the shift

\begin{equation}
\label{eq:algoriths 6 and 7 equivalence}
    \begin{split}
        \xi_1^k &= x_3^k \\
        \xi_2^k &= x_1^{k + 1},
    \end{split}
\end{equation}
which relates the initial conditions of Algorithm \ref{algo : 7} ($\xi_i^0$) to iterates of Algorithm \ref{algo : 6} ($x_i^k$), for $k \geq 0$. After performing such a shift, the two algorithms make the same calls on the proximal operators. 

\begin{algorithm}
\caption{}
\label{algo : 6}
\begin{algorithmic}
\For{$k = 0, 1, 2, ..., K$} 
\State $x_1^{k + 1} = \text{prox}_f(x_3^k)$
\State $x_2^{k + 1} = \text{prox}_g (2x_1^{k + 1} - x_3^k)$
\State $x_3^{k + 1} = x_3^k + x_2^{k + 1} - x_1^{k+1}$
\EndFor
\end{algorithmic}
\end{algorithm}

\begin{algorithm}
\caption{}
\label{algo : 7}
\begin{algorithmic}
\For{$k = 0, 1, 2, ..., K$} 
\State $\xi_1^{k + 1} = \text{prox}_g (-\xi_1^{k} + 2 \xi_2^k) + \xi_1^k - \xi_2^k$
\State $\xi_2^{k + 1} = \text{prox}_f(\xi_1^{k + 1})$
\EndFor
\end{algorithmic}
\end{algorithm}

\begin{figure}
    \centering
    \includegraphics[width = 0.47\textwidth]{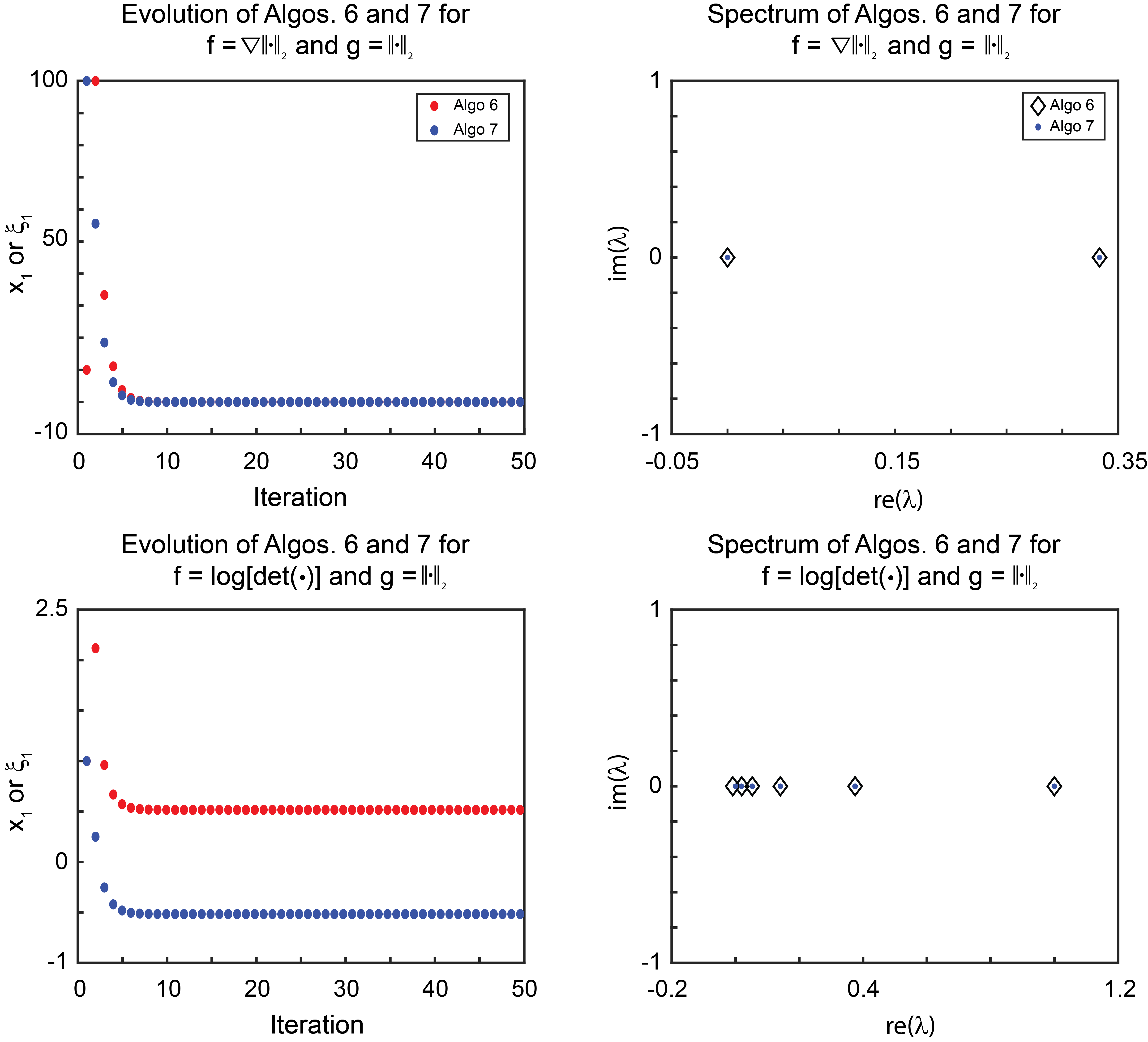}
    \caption{\textbf{Shift equivalence of Algorithms \ref{algo : 6} and \ref{algo : 7}.} Left column, trajectories of $x^k_1$ and $\xi^k_1$ for choices of $f = \nabla || \cdot ||_2$ and $g = || \cdot ||_2$ (top row) and $f = \log[\det(\cdot)]$ and  $g = || \cdot ||_2$ (bottom row). Right column, corresponding Koopman spectra. In both cases, the overlapping spectra correctly identify the ``shift'' equivalence \cite{zhao2021automatic}.}
    \label{fig:Shift Equivalence}
\end{figure}

Looking at the sequence of outputs, $x^k_1$ and $\xi^k_1$, we see that indeed, ignoring $x_1^0$, the two have the same dynamics, for different choices of $f$ and $g$ (Fig. \ref{fig:Shift Equivalence} left column). Therefore, we expect that the Koopman eigenvalues will be similar between the two shifted systems. Computing the Koopman spectra of Algorithms \ref{algo : 6} and \ref{algo : 7}, we confirm that it is indeed possible to identify this kind of equivalence with no changes to our framework (Fig. \ref{fig:Shift Equivalence}, right column), as the Koopman spectra are largely overlapping. 

Note that, as long as a sufficient amount of data snapshots are present, we can expect the two Koopman spectra to converge \cite{klus2016numerical, korda2018convergence}. An insufficient amount of data could, however, lead to differing decompositions. This could be overcome by using methods designed for the sparse data regime \cite{sinha2019sparse, sinha2021shot}.

\section{Discussion}
\label{sec:discussion}
By viewing iterative algorithms applied to a given problem as discrete-time dynamical systems \cite{dietrich2020algorithms}, we developed a framework for identifying equivalent algorithms via the spectra of the associated Koopman operators. The key to this approach relies on the fact that two dissipative systems that are conjugate have the same Koopman eigenvalues \cite{mezic2020spectrum}. Similarly, two dissipative systems that are semi-conjugate have Koopman eigenvalues, where one is a subset of the other.

That we were able to make comparisons between dynamical systems in an applied setting using these mathematical results illustrates the wide implications that the rigorous theoretical development of Koopman operator theory can have. Indeed, results connecting Koopman spectral objects to geometric properties of state space are indispensable for gaining insight when applying Koopman tools \cite{mezic2005spectral,  mauroy2013iso, lan2013linearization,  arbabi2017ergodic, mezic2020spectrum}.

Our method for identifying algorithmic equivalence necessarily requires the computation of the Koopman eigenvalues from data, which makes it necessary to have a sufficient amount of data and an appropriate choice of numerical scheme for a good approximation. Its conclusion of equivalence can depend on the domain chosen, as in the case of algorithms that are locally conjugate. This is in contrast to the linear control based method developed by Zhao et al. (2021), which was analytic and enabled identification of equivalences from the equations defining the algorithms, independent of initial conditions \cite{zhao2021automatic}.

However, our framework avoids several of the limitations of the linear control framework \cite{zhao2021automatic, lessard2016analysis}. First, it was possible to identify nonlinear equivalences between algorithms. Second, by considering just the Koopman eigenvalues, it was possible to identify (semi-)conjugacy, unifying the several different definitions of equivalence introduced by Zhao et al. (2021). Third, it was possible to identify equivalence without the underlying equations. As long as a sequence of outputs from the algorithms are available, the Koopman framework can be used, making it especially useful in the case of proprietary software. Fourth, it can be used to distinguish between conjuagcies with different properties (such as local and global conjugacy), an important distinction when comparing algorithms. Finally, this was all possible solely by viewing algorithms as discrete-time dynamical systems. No additional considerations, such as control, were required. 

Taken together, these result illustrate that Koopman operator theory is a useful framework with which to study algorithmic equivalence, as well as algorithms more generally.  



\section*{Acknowledgement} 

W.T.R. was partially supported by a UC Chancellor’s Fellowship. W.T.R., M.F., R.M and I.M were partially supported by the Air Force Office of Scientific Research project FA9550-17-C-0012. I. G. K. was partially supported by the U.S. DOE (ASCR) and DARPA ATLAS programs.


\small
\bibliography{main.bib}
\bibliographystyle{unsrt}

\end{document}